\documentclass[11pt,titlepage]{article}

\usepackage{setspace} 

\usepackage{amssymb,amsfonts,amsmath}

\usepackage{graphicx}

\usepackage[round,numbers,sort&compress]{natbib} 
\title{Switching of swimming modes in \textit{{Magnetospirillium gryphiswaldense}}}

\author{M. Reufer *\thanks{Corresponding author.  email: mathias.reufer@gmx.ch}, R. Besseling *, J. Schwarz-Linek *, \\V. A. Martinez *, A. N. Morozov *, J. Arlt *, D. Trubitsyn$^\dagger$,\\ F. B. Ward$^\dagger$, W.~C.~K.~Poon *\\
\\
*SUPA and COSMIC, School of Physics \& Astronomy, \\
The University of Edinburgh, King's Buildings,\\ Edinburgh EH9 3JZ, United Kingdom 
\and $^\dagger$Institute of Cell Biology, The University of Edinburgh,\\ King's Buildings, Edinburgh EH9 3JR, United Kingdom}

\date{\today}

\begin{document}

\maketitle

\abstract{
The microaerophilic magnetotactic bacterium \emph{Magnetospirillum gryphiswaldense} swims along magnetic field lines using a single flagellum at each cell pole. It is believed that this magnetotactic behavior  enables cells to seek optimal oxygen concentration with maximal efficiency. We analyse the trajectories of swimming \emph{M. gryphiswaldense} cells in external magnetic fields larger than the earth's field, and show that each cell can switch very rapidly (in $< 0.2$~s) between a fast and a slow swimming mode. Close to a glass surface, a variety of trajectories was observed, from straight swimming that systematically deviates from field lines to various helices. A model in which fast (slow) swimming is solely due to the rotation of the trailing (leading) flagellum can account for these observations. We determined the magnetic moment of this bacterium using a new method, and obtained a value of $\left( 2.0 \pm 0.6 \right) \times 10^{-16}$~A$\cdot$m$^2$. This value is found to be consistent with parameters emerging from quantitative fitting of trajectories to our model. \\}


\clearpage
\section*{Introduction}
Cells of magnetotactic bacteria \citep{BlackmoreReview1982} possess iron-bearing magnetosomes with permanent magnetic moments, so that they passively align parallel to and swim along magnetic field lines: they are `flagella-propelled compass needles'. The adaptive value of magnetotaxis is not fully understood. Since all known species live in stratified water columns or sediments, and many are obligate microaerophiles, it is likely that magnetotaxis works in tandem with oxygen-seeking behavior -- magnetoaerotaxis -- to guide organisms to preferred redox environments. Aligning to geomagnetic field lines provides an inexpensive (because passive) way of sensing of `up' and `down', which is the predominant direction of redox gradients. 

Two types of magnetoaerotaxis have been identified \citep{FrankelReview2006}, but not yet fully understood. In {\it polar} magneto-aerotaxis, cells swim towards the geomagnetic north (south) in the northern (southern) hemisphere in oxidizing conditions, usually propelled by single polar flagella. In reducing conditions, organisms reverse their  flagellar motors, and hence their velocity. Many live at oxic-anoxic interfaces, where oxidizing/reducing conditions prevail above/below respectively. By contrast, {\it Magnetospirilla} are {\it axial} magnetoaerotactic. In a uniform environment, they swim towards both geomagnetic poles. Each cell senses the oxygen concentration along its trajectory as a function of time. If a cell detects an unfavorable gradient (= motion away from preferred conditions), it increases the probability of reversing its flagellar motor, and therefore velocity. The opposite happens along a favourable gradient.

Such temporal gradient sensing is reminiscent of chemotaxis in {\it Escherichia coli} \citep{BergBook}, where a rod-shaped cell body is propelled by a rotating bundle of rigid helical flagella \citep{BergBook}. Propulsion in {\it Magnetospirilla} is quite different. {\it Magnetospirilla} are bipolarly flagellated, bearing one flagellum that is {\it not} helical in its quiescent state at each pole of a mature cell, which {\it is} helical \citep{Schuler1992}. Propulsion is likely similar to that in the non-magnetic species {\it Spirillum voluntas}, where each polar `flagellum' is actually a bundle of about 75 thinner flagella \citep{Swan1982}. Dark-field observations \citep{Metzner1923,Krieg1967} show that both flagella bundles in a mature cell rotate in the same direction during swimming. The simplest propulsion model \citep{Chwang1972} treats the cell body as a rigid helix, and the polar flagella as rigid rods. The rotating flagella cause the cell body to rotate in the opposite direction to render the whole organism torque free. The rotational-translational coupling of the body spiral then propels it forward. More sophisticated models take account of the chiral shapes adopted by the rotating flagella \citep{Winet1976,Ramia1991,Ramia1994}.

Understanding magnetotactic bacteria motility is interesting and important from many perspectives. The ecological niches they inhabit, which are intimately linked to their magnetoaerotaxis, give them a potentially important but as yet little understood role in global iron cycling \citep{Simmons}. In the biophysics of micro-propulsion \citep{LaugaPowers}, these organisms provide novel paradigms beyond the well-studied {\it E. coli}. Since motile bacteria increasingly function as models for the physics of `active colloids' \citep{JanaPNAS}, magnetotactic organisms open up the possibility of studying such systems in an easily tunable external field. For applications, magnetotactic bacteria `guided' by external fields have been used to manipulate mesoscopic `cargoes' \citep{Martel2006}, and proposed as  drug delivery vehicles \citep{Martel2007}. 

We study the swimming of  {\it M. gryphiswaldense}, one of the better-characterized magnetotactic bacteria. Sampling of $\lesssim 300$ organisms returned a broad speed distribution between 5-50~$\mu\mbox{ms}^{-1}$ \citep{Martel2006}.  Comparative genomics has pinned down `core genes' for magnetotaxis \citep{genome}. The kinetics of its magnetosome formation has been followed using X ray scattering \citep{Ward2007}. It is not yet possible to visualize the individual flagella of {\it M. gryphiswaldense} -- significantly thinner than the bundles in {\it S. voluntas} -- during swimming.

In a rotating magnetic field, {\it M. gryphiswaldense} shows a variety of complex trajectories \citep{Erglis2007}; the same study also found incidentally that the organism could reverse its velocity during motion, though the mechanism is unknown. Perhaps surprisingly, there has been no quantitative study in the simpler environment of {\it constant} external magnetic field to date. In this paper, we report such a study, which yields an unexpected result: the speed distribution is strongly bimodal. Cells swimming in either direction at $\approx 15~\mu\mbox{ms}^{-1}$ or $\approx 45~\mu\mbox{ms}^{-1}$, and can switch rapidly (in $\lesssim 0.2$~s) between the fast and slow modes. To elucidate the biomechanical origins of this behavior, we track cells next to a glass slide in an external magnetic field. Wall-cell hydrodynamic interaction competes with magnetic effects to generate a variety of trajectories, all of which can be accounted for by the hypothesis that the fast and slow swimming modes are associated with deploying the trailing and leading flagellum respectively. It is possible that such switching may be part of the organism's strategy for exploring its environment {\it perpendicular} to magnetic field lines.

\section*{Materials and Methods}

\subsection*{Bacteria Strain and Sample Environment} Liquid cultures of {\em M. gryphiswaldense}, strain MSR-1,  were grown at room temperature (approximately 22$\,^{\circ}$C) following established protocol \citep{Schultheiss2003} in closed 25~ml vials with 10~mm air above the surface over 3 to 5 days. We selected motile cells with magnetosomes by loading a sterile 5~ml syringe with the culture. The syringe was placed for about 4 h next to a permanent magnet (with S-pole towards the tip) so that motile cells would swim along field lines in a field of $\approx 5$ mT and accumulate close to the tip.

\subsection*{Magnetic field} The $z$-field was applied using a 20 mm diameter electromagnet mounted just below the sample. The $y$-field was due to two disc-shaped permanent magnets mounted at $y=\pm$~20 mm to give $B_y=-1.5 \pm 0.1 $~mT in a central region of $10 \times10 $~mm$^2$ where the sample was placed.

\subsection*{Microscopy and Image Analysis} Image series containing approximately 3700~images were collected at 100~frames per second using a Nikon Eclipse Ti inverted microscope equipped with CMOS-camera (Mikrotron MC 1362) and frame grabber card (Mikrotron Inspecta 5). The bacterial suspension was filled into a custom-made observation chamber consisting of a cover-glass with spacer (160~$\mu$m thick cover-glass) glued with UV-glue (Loctite 358) on a microscopy slide and sealed with Vaseline. We used a phase contrast objective (Nikon 10x Ph1) with $\approx 30-40~\mu$m depth of field, which enabled us to view selectively either bacteria swimming along the bottom or the top of the observation chamber. We used a $\times 100$ objective to estimate the swimming distance of cells from the bottom surface.

We collected 30~s image sequences each with approximately 500 swimming cells in the field of view. Trajectories were extracted  using standard tracking methods \citep{Crocker1996}. We discarded trajectories that showed diffusive rather than ballistic (swimming) behavior using the method of Mi\~{n}o et al. \citep{Mino2011}. To achieve better statistics we used up to 15 image sequences for constructing swimming speed distributions.

\subsection*{Magnetic moment determination}

We used differential dynamic microscopy (DDM) recently applied to studying swimming of microorganisms \citep{LaurenceDDM,VincentDDM} and generalised to anisotropic particles \citep{MathiasDDM} to determine the effective translational diffusion coefficients of non-motile {\em M. gryphiswaldense} parallel and perpendicular to an applied horizontal magnetic field. Non-motile cells were obtained by heating wild-type organisms at 60$^\circ$C for 1 hour.
In DDM, we analyse the intensity fluctuations from the spatial Fourier transform of a sequence of images to obtain the intermediate scattering function (ISF), $f(\mathbf{q},t)$, where $\mathbf{q}$ is the wave vector of the fluctuations being probed, and $t$ is time. The ISF measures dynamics on the spatial scale $\sim 2\pi/q$. By analysing $f(\mathbf{q},t)$ in two sectors with $\mathbf{q}$ approximately perpendicular and parallel to the applied field $\mathbf{B}$, we can determine the effective translational diffusion coefficients of the particles in a dilute suspension as a function of $B$. Fitting the observed field-dependence of these diffusivities gives a value for $m$. 

\begin{figure}
\begin{center}
\includegraphics[width=3.25in]{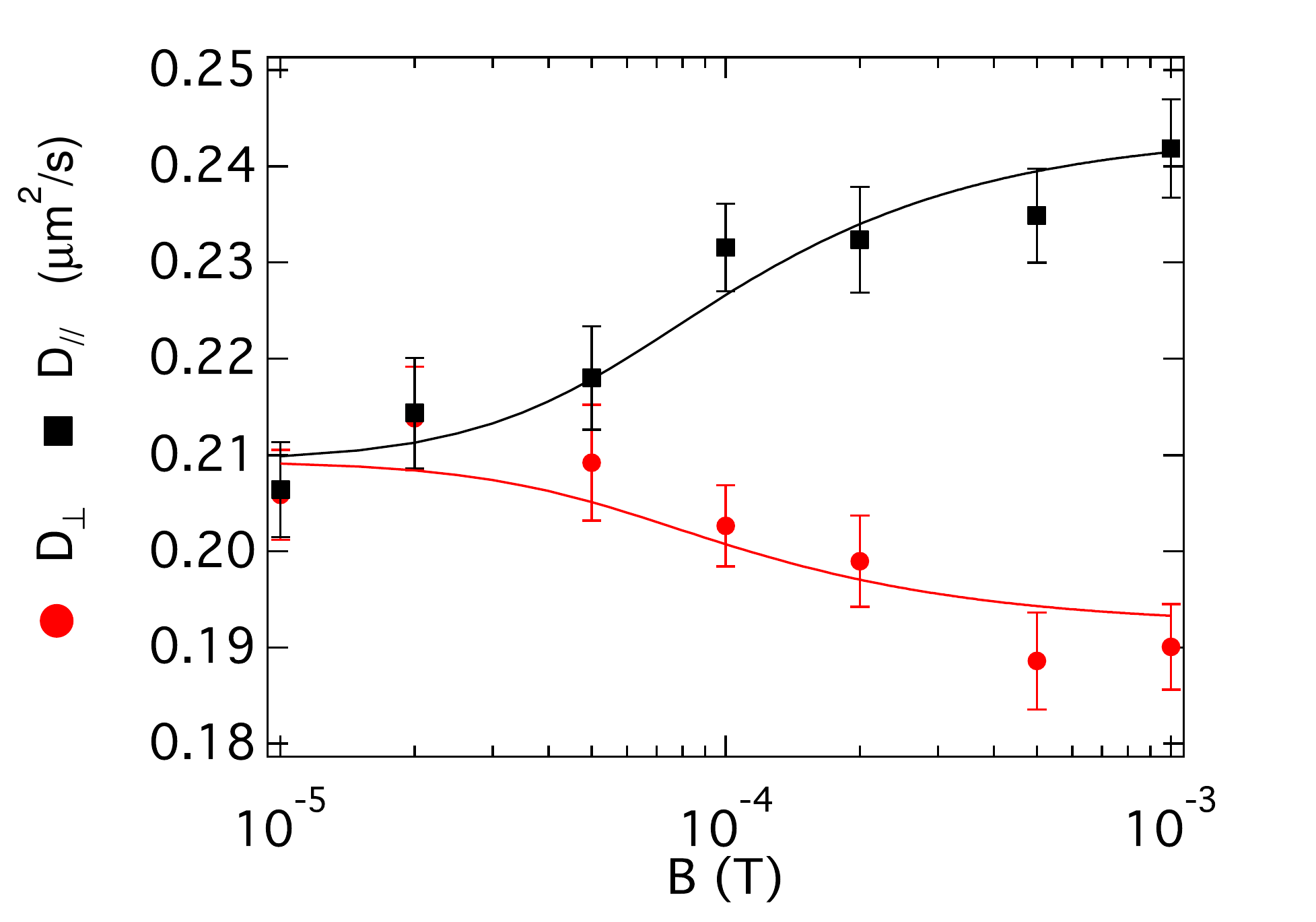}
\caption{}
\label{fig:D_vs_B}
\end{center}
\end{figure}

In a mean-field kinetic theory, the effective translational diffusivities of non-interacting ellipsoidal particles (each with fixed magnetic moment $\mathbf{m}$) perpendicular, $D_{\perp}$, and parallel, $D_{\parallel}$, to an applied magnetic field, $\mathbf{B}$, are given by \citep{Ilg_2005}:
\begin{eqnarray}
D_{\perp} & = & D_{\rm iso} -\frac{1}{3}(D_a-D_b)S_2(h), \\     
D_{\parallel} & = & D_{\rm iso} +\frac{2}{3}(D_a-D_b)S_2(h),    
\end{eqnarray}
where $h=\frac{mB}{k_B T}$ is the reduced magnetic field, $k_B$ is the Boltzmann's constant, $D_{\rm iso}=\frac{D_a+2D_b}{3}$ is the isotropic
translational diffusion coefficient,  and $D_a$, $D_b$ are the
translational diffusivities of a prolate spheroid along the major axis and minor axis respectively at $B = 0$. In the case of particles with $\mathbf{m}$ along the major axis, the second-order orientational order parameter is given by
\begin{equation}
S_2(h)= 1-3 \left[\frac{ \text{coth}(h)-h^{-1} }{h}\right].
\end{equation} 
This quantifies the mean alignment of the particles with increasing $B$ and therefore the transition from isotropic to anisotropic dynamics. In the zero and infinity field limits, $S_2(h)$ reduced to 0 and 1 respectively, giving:
\begin{eqnarray}
& D_{\parallel}(B=0)=D_{\perp}(B=0)=D_{\rm iso}\\
& D_{\parallel}(B_{\infty})=D_a\\
& D_{\perp}(B_{\infty})=D_b
\end{eqnarray}
$D_{\perp}$ and $D_{\parallel}$ were measured using DDM over a range of spatial frequency $1\leq q \leq 2.3~\mu$m$^{-1}$ as functions of $B$, Fig.~\ref{fig:D_vs_B}. Fitting to mean-field kinetic theory expressions given above gives estimates of $D_a$, $D_b$ and $m$ (see caption of Fig.~\ref{fig:D_vs_B}) .

The diffusivities $D_a$ and $D_b$ are related to the corresponding drag coefficients of a prolate spheroid:
\begin{equation}
D_a = \frac{k_B T}{\xi_a} \qquad \text{and} \qquad D_b = \frac{k_B
  T}{\xi_b}, 
\end{equation}
where \citep{Chwang1975}
\begin{eqnarray}
&&\xi_a = \frac{16\pi \eta a
  \epsilon^3}{\left(1+\epsilon^2\right)\log{\frac{1+\epsilon}{1-\epsilon}}
- 2\epsilon}, \\
&&\xi_b = \frac{32\pi \eta a
  \epsilon^3}{\left(3\epsilon^2-1\right)\log{\frac{1+\epsilon}{1-\epsilon}}
+ 2\epsilon}.
\end{eqnarray}
Here $\eta$ is the viscosity of the surrounding fluid and $\epsilon=\sqrt{1-\frac{b^2}{a^2}}$, with $a$ and $b$ the major and minor semi-axis, respectively. Solving numerically Eqs. 8 $\&$ 9, we found $a\approx2.1$~$\mu$m and
$b\approx0.6$~$\mu$m, using the viscosity of water $\eta=10^{-3}$~Pa$\cdot$s and the measured diffusion coefficient $D_a=0.244$~$\mu$m$^2/$s and  $D_b=0.192$~$\mu$m$^2/$s.

\section*{Results}

We cultured cells of {\it M. gryphiswaldense} according to published protocol \citep{Schultheiss2003} and selected organisms with magnetosomes using a standard `race track' method. These were then loaded into 160~$\mu$m thick sealed sample chambers and observed at $\times 10$ magnification in a phase contrast microscope. Movies were taken using a CCD camera, typically at 100 frames per second. (See Materials and Methods for details.) 

\subsection*{Bimodal Speed Distribution}

\begin{figure}
\begin{center}
\includegraphics[width=3.25in]{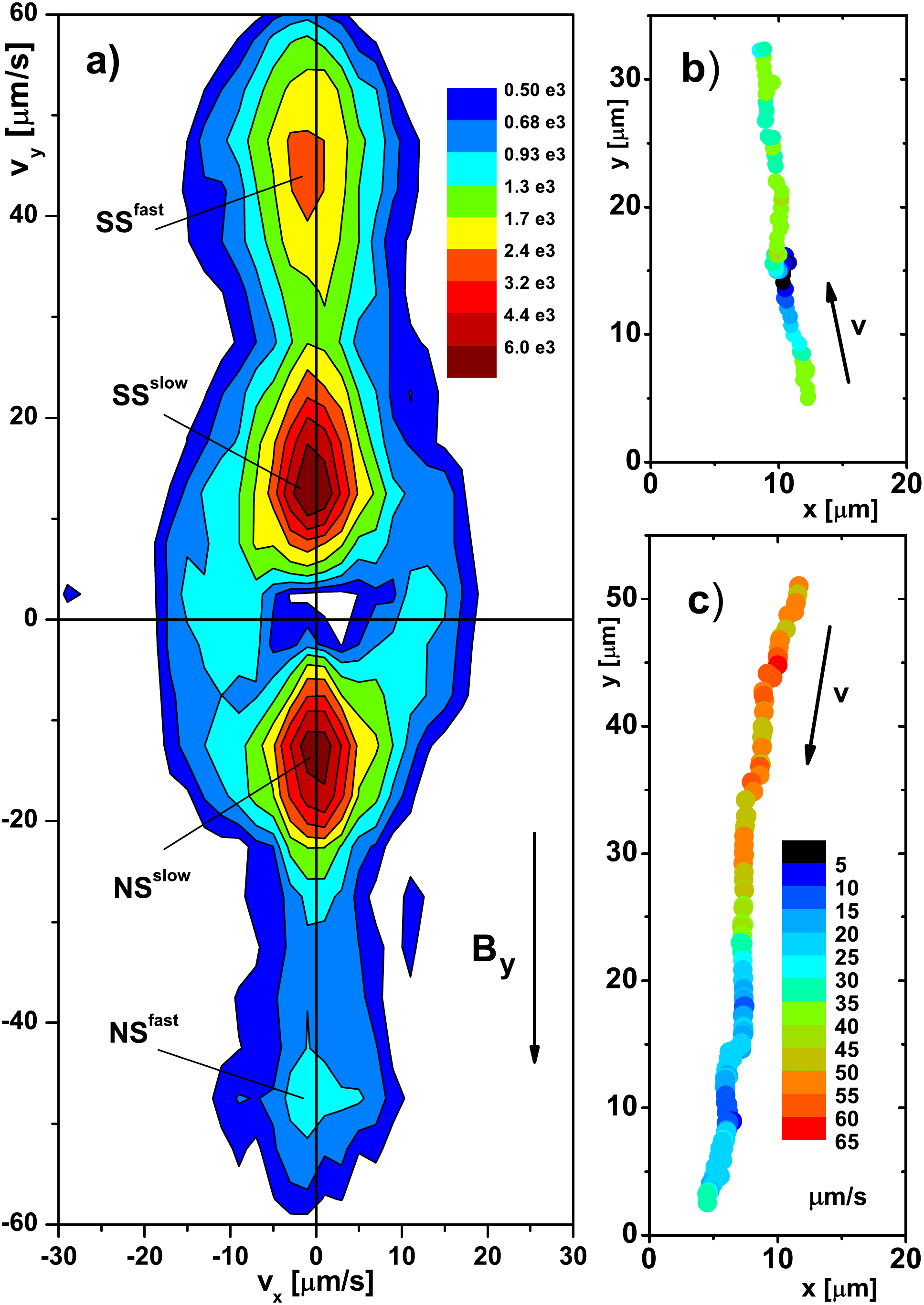}
\caption{}
\label{Polar_Bulk}
\end{center}
\end{figure}

We first investigated swimming in the bulk, i.e. far away from surfaces, under a homogeneous horizontal applied magnetic field $B_y=-1.5$~mT ($50 \times$ earth's field in magnitude), $B_x = B_z = 0$. Here and throughout, the applied field is antiparallel to the $y$ axis, which, together with an orthogonal $x$ axis define the plane that is being imaged. The optic axis of our microscope defines the $z$ axis in a right-handed system, Fig.~\ref{MicroCell}. Approximately equal numbers of the cells swim towards the magnetic north pole (parallel to $y$) and towards the magnetic south pole (antiparallel to $y$).\footnote{More precisely, 44\% of 500 trajectories were NS.} These are south-seeking (SS) and north-seeking (NS) respectively, since the earth's magnetic south pole is situated near the earth's geographic North Pole. Fig.~\ref{Polar_Bulk}(a) shows the distribution of swimming velocities in the $xy$ plane, $P(\mathbf{v})$, calculated over 0.4~s sections of trajectories. For both SS and NS organisms, the distribution is clearly bimodal with peaks corresponding to speeds of 15~$\mu$m/s and 45~$\mu$m/s. We will refer to these peaks as SS$^{\rm slow}$, SS$^{\rm fast}$, NS$^{\rm slow}$, and NS$^{\rm fast}$.

Examination of individual trajectories shows that this distribution results not from two distinct populations, but from  cells varying their swimming speeds. Fig.~\ref{Polar_Bulk}(b) shows the trajectory of a SS$^{\rm fast}$ cell that stopped for 0.7 s before continuing SS$^{\rm fast}$; Fig. \ref{Polar_Bulk}(c) shows the trajectory of a NS organism that switched from fast to slow swimming. We found trajectories with all possible speed and velocity changes, although speed switching with no reversal is the most common.

\subsection*{Near-wall Swimmers Deviate From Magnetic Field Lines} 
\begin{figure}
\begin{center}
\includegraphics[width=3.25in]{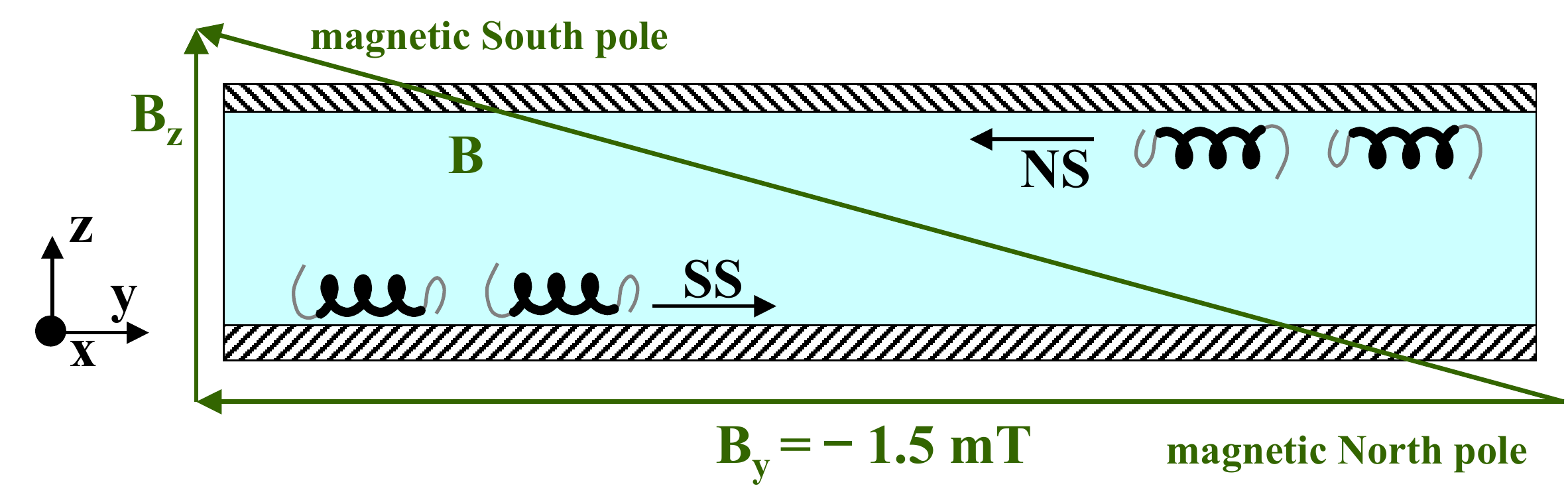}
\caption{}
\label{MicroCell}
\end{center}
\end{figure}

\begin{figure}
\begin{center}
\includegraphics[width=3.25in]{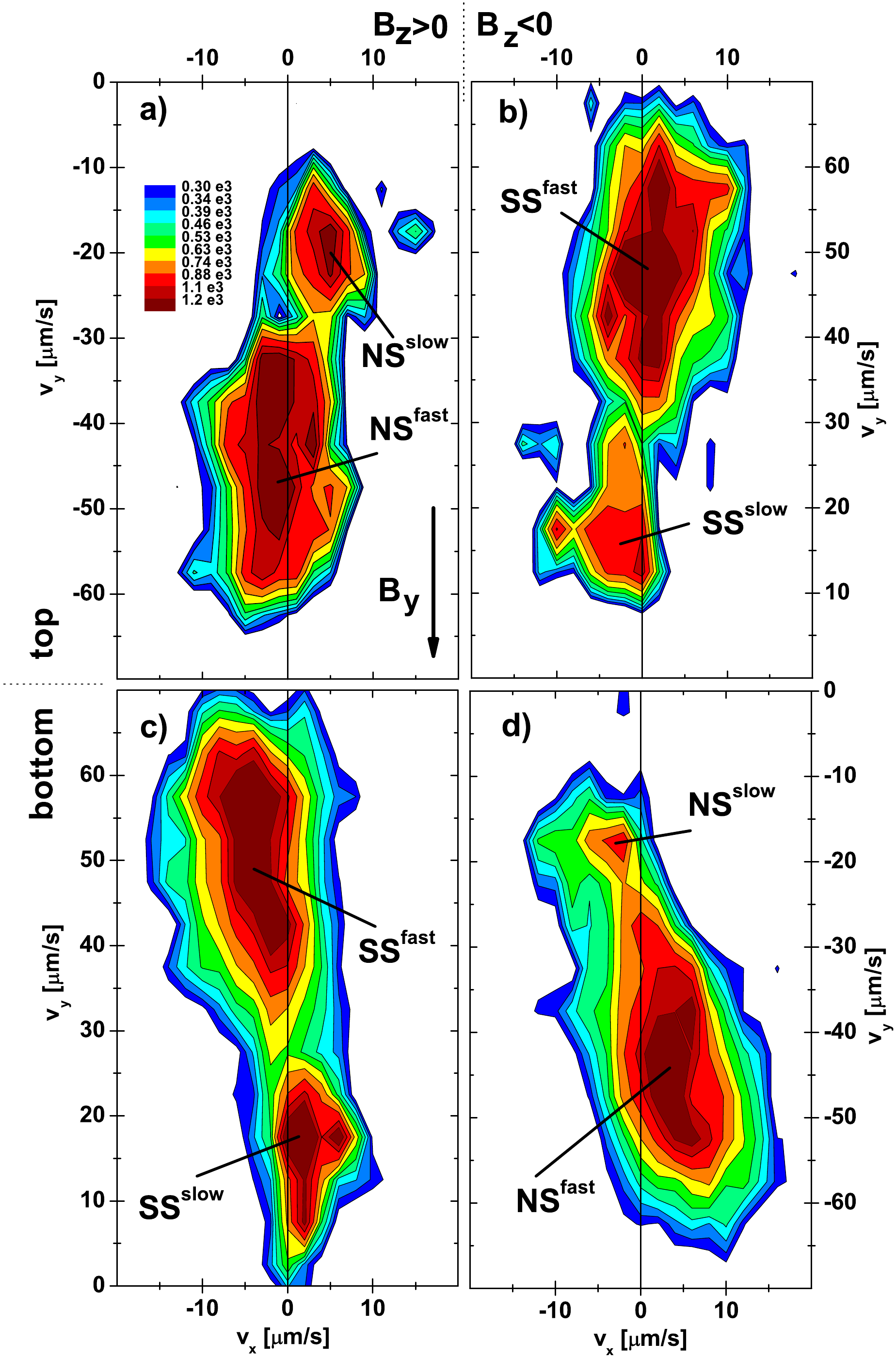}
\caption{}
\label{Spot}
\end{center}
\end{figure}

\begin{figure}
\begin{center}
\includegraphics[width=3.25in]{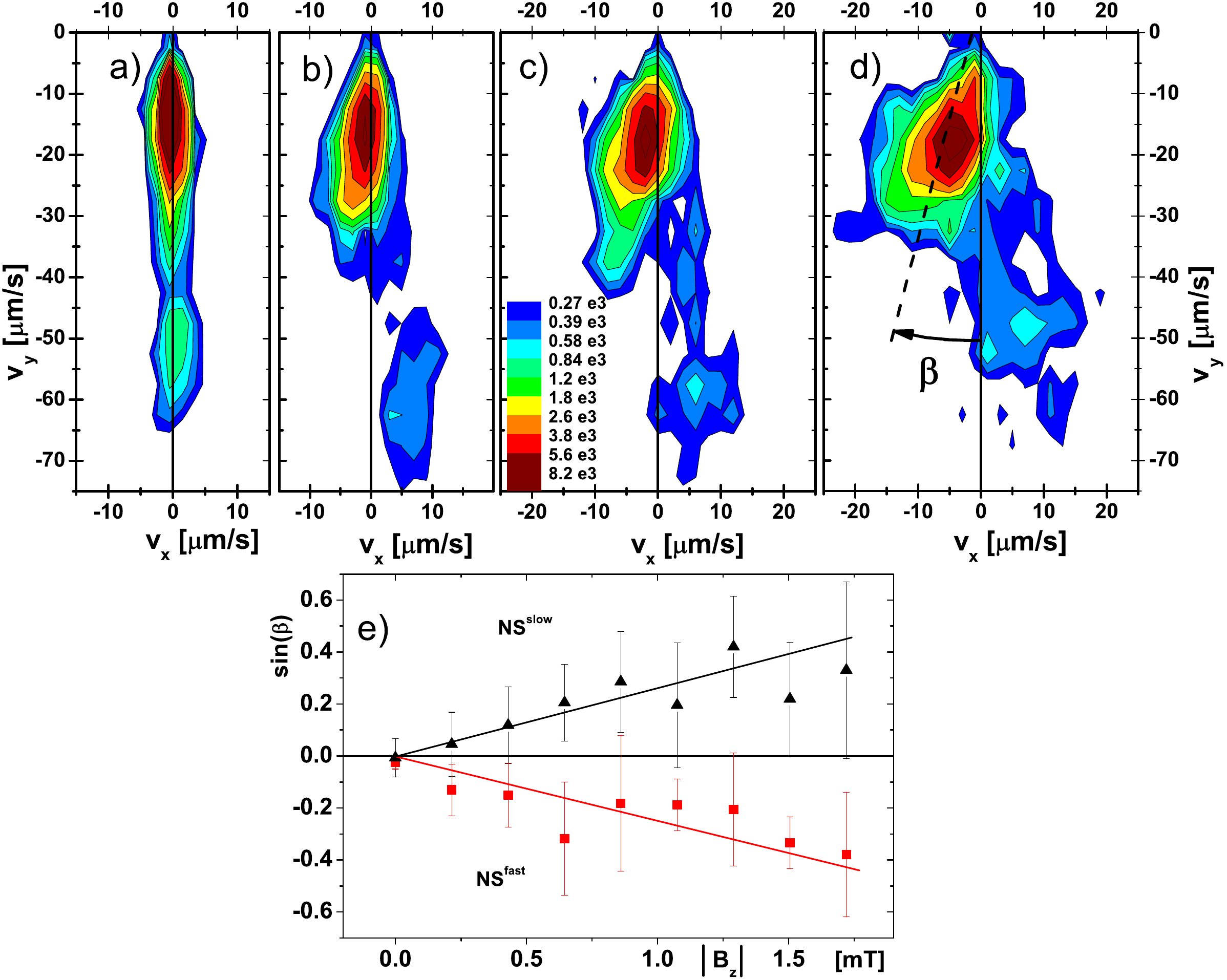}
\caption{}
\label{FieldDependency}
\end{center}
\end{figure}

We next added a vertical component, $B_z$, to the applied field so that NS (SS) cells are `guided' by the now slant field lines to, and accumulate at, the top (bottom) for $B_z>0$; for $B_z<0$ the accumulation is on the opposite side, Fig.~\ref{MicroCell}. Fig.~\ref{Spot} shows $P(\mathbf{v})$ of these near-wall swimmers at $B_z = \pm 0.86$~mT. Since the separation of NS and SS organisms was total, only the relevant half of $P(\mathbf{v})$ is shown. In contrast to Fig.~\ref{Polar_Bulk}, where on average $\mathbf{v} \parallel \pm \mathbf{B}$, there is now a clear deviation of $\mathbf{v}$, the projected $xy$ velocity, from the direction of the field lines in the $xy$ plane, which are along $y$. The angular deviation, $\beta$, has opposite signs for fast and slow swimmers, Fig.~\ref{Spot}. Its mean magnitude, $\bar\beta$, increases with the vertical field, Fig.~\ref{FieldDependency}(a)-(d), so that $\sin\bar\beta \propto |B_z|$ approximately, Fig. \ref{FieldDependency}(e). Importantly, the bulk result ($\bar\beta = 0$, Fig.~\ref{Polar_Bulk}) is recovered at a wall at $B_z = 0$. 
Observations at a higher numerical aperture revealed that cells swam at $3\pm2$ $\mu$m from the wall at $|B_z|>0.5$ mT, increasing to $8\pm6$ $\mu$m at $| B_z | \approx0.1$ mT.

We discuss the origins of these deviations later. Here we simply point out that they do not arise from a misalignment of $B_y$ or $B_z$ such that $B_x \neq 0$, because the observed sense of deviation is opposite for cells swimming along the top and the bottom of the sample chamber under the same conditions (i.e. comparing Fig.~\ref{Spot}(a) with \ref{Spot}(c) or comparing \ref{Spot}(b) with \ref{Spot}(d)).

Interestingly, there is no detectable change in the absolute swimming speeds near a wall from their values in the bulk: in both cases, the majority of cells swim at $\approx 45~\mu$ms$^{-1}$ and $\approx 15~\mu$ms$^{-1}$, Figs.~\ref{Polar_Bulk} and \ref{Spot}. This means that while the field lines guiding cells to the wall are at an angle to horizontal ($\alpha = \tan^{-1}( B_z/B_y)$, Fig.~\ref{MicroCell}), cells swim essentially parallel to the top and bottom surfaces. The origins of the torque (in the $-x$ direction) giving rise to thFis state of affairs is unclear, but a similar torque has been observed before for {\it Caulobacter crescents} cells swimming next to a surface   \citep{Tang2009}.

\subsection*{Observing surface mode switching and loops} 
\begin{figure}
\begin{center}
\includegraphics[width=3.25in]{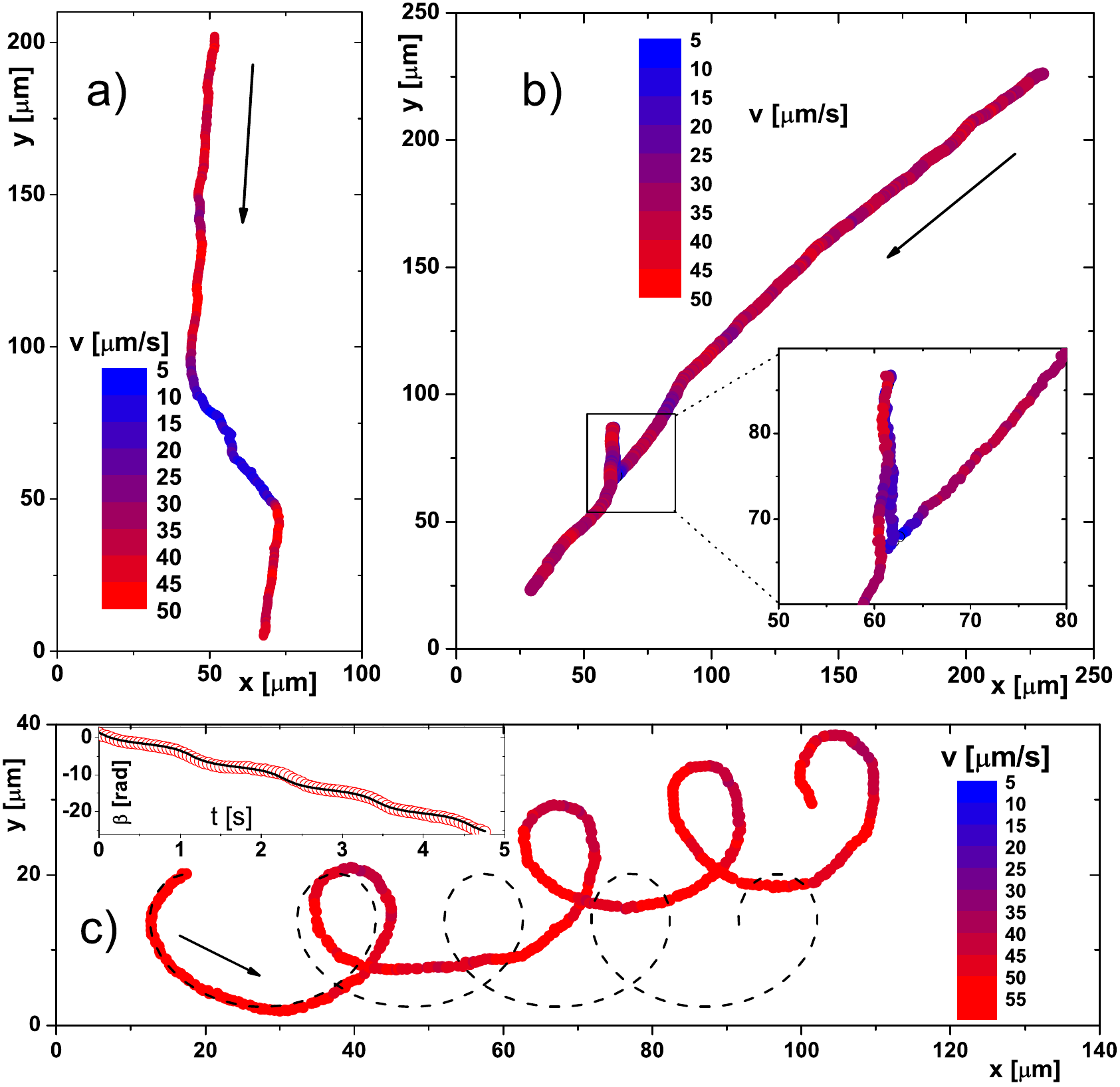}
\caption{}
\label{InterestingTracks}
\end{center}
\end{figure}

Fig. \ref{InterestingTracks}(a) shows the trajectory of a cell switching from NS$^{\rm fast}$ (red) to NS$^{\rm slow}$ (blue) and back to NS$^{\rm fast}$ (red), with deviations consistent with Fig.~\ref{Spot}(a);  Fig.~\ref{InterestingTracks}(b) shows a NS$^{\rm fast}$ cell reversing to SS$^{\rm slow}$, and then reverting to NS$^{\rm fast}$.
Interestingly, $\beta$ changes abruptly at the first reversal from $\approx~45\,^{\circ}$ to  $\approx~0\,^{\circ}$. The change from NS to SS means that the cell leaves the top surface, Fig.~\ref{MicroCell}, and starts swimming into the bulk, where no deviation occurs, Fig.~\ref{Polar_Bulk}. After reversing to NS, the cell returns to the wall and shows $\beta~\approx~45\,^{\circ}$ again. In 370~such trajectories, each of  duration $\gtrsim 3.5$~s, $\approx 25~$\% contained swimming mode changes. All possible combinations of mode switching between SS$^{\rm slow}$, SS$^{\rm fast}$, NS$^{\rm slow}$, and NS$^{\rm fast}$ were observed.  Each putative switching event was checked to ensure that it was not an artefact due to intersecting trajectories. 

A very small proportion ($\approx 0.2\%$, found by automated scanning through $10^4$ trajectories) of the trajectories at  $| B_z |>$~1~mT contained loops, Fig.~\ref{InterestingTracks}(c). Loops were not observed at lower vertical fields. The sense of these loops depends on the swimming speed. At $v<22~\mu\mbox{ms}^{-1}$ we find CW loops on the bottom and CCW ones on the top of the sample cell (viewed from the top), whereas for $v >36~\mu\mbox{ms}^{-1}$ the sense of the loops are reversed. A mixture of these kinds of behavior was obtained at $22 < v < 36~\mu\mbox{ms}^{-1}$.

\section*{Modeling two-speed swimming}
Our observations show that {\it M. gryphiswaldense} in an external magnetic field switches between four swimming modes:  NS$^{\rm slow}$, NS$^{\rm fast}$, SS$^{\rm slow}$ and SS$^{\rm fast}$. Two velocity modes have been observed before in {\it S. voluntas}. In a young cell, which possesses only a single flagella bundle immediately after cell division, propulsion by the trailing bundle is twice as fast as propulsion by the leading bundle \citep{Swan1982}. In a previous study of {\it M. gryphiswaldense} \citep{Erglis2007}, an asymmetry of the swimming speed between NS and SS was reported. It was suggested that this asymmetry was due to a change of the flagella pitch upon reversal. In this work, we find that reversal and the switch between fast and slow swimming modes are independent phenomena, e.g.~changing from NS$^{\rm fast}$ to NS$^{\rm slow}$ and back to NS$^{\rm fast}$ without reversal is possible, Fig.~\ref{InterestingTracks}(a). We therefore seek another explanation for two-speed swimming. 

\subsection*{Wall Hydrodynamic Interactions} An important clue comes from the observation that cells next to a wall swim at a finite deviation to the direction of the horizontal magnetic field. Thus, each cell experiences a finite magnetic torque, which must be balanced by another torque; the latter is due to hydrodynamic interactions between cells and the wall.

 In a rotating body close to a solid wall  \citep{BrennerBook}, the drag coefficient of any element of the body increases with decreasing distance from the wall, resulting in a net force perpendicular to the angular velocity and the wall normal. The direction of this force reverses if the sense of rotation reverses. Thus, {\it E. coli} swims in circles next to solid surfaces \citep{Lauga2006}: the body and the flagella bundle rotate in opposite directions, giving rise to equal and opposite hydrodynamic forces that act on the body and the flagella bundle respectively,  so that the whole cell experiences a net hydrodynamic torque, which is balanced by the viscous torque due to the cell's rotation in a viscous fluid.

\subsection*{Quantitative Model} {\it M. gryphiswaldense} possesses a helical body with two polar flagella. We assume that each rotating flagellum adopts a spiral form, as is observed in {\it S. voluntas}. Our proposed physical picture is as follows. Suppose a cell swims by rotating both flagella. Then, borrowing directly from the analysis of {\it E. coli} next to surfaces \citep{Lauga2006}, the system of hydrodynamic forces acting on a cell swimming next to a bottom wall is as shown in Fig.~\ref{model}(a), with the (force-free) requirement that $F_{\rm body}=F_{\rm lead}+F_{\rm trail}$. If, to a first approximation, we take the front and back flagella to be  equivalent, then $F_{\rm lead} = F_{\rm trail}$,  and the net hydrodynamic torque is zero, $T_{\rm hyd} = 0$. With no need for any balancing magnetic torque, $mB_y\sin\beta = 0$, or $\beta = 0$, i.e. cells should swim along field lines. This is not what is observed. 

We therefore hypothesise that cells swim with either the leading or trailing flagella, which does give rise to net hydrodynamic torques along the $z$ axis with opposite signs, Fig.~\ref{model}(a). If the magnetic field is large enough, viz., $| T_{\rm hyd} | < | m  B_y |$, a solution can be found for $T_{\rm hyd}=- m  B_y \,{\rm sin}\beta$. Cells will swim at a fixed $\beta \neq 0$ to the field lines, with the sign of $\beta$ dependent on the sign of $T_{\rm hyd}$, which in turn is a function of whether the leading or trailing flagellum is in use. Rotation of the leading flagella alone results in a deviation of swimming direction in CW sense with respect to the bulk swimming direction, while rotation of the trailing flagella alone leads to a deviation in CCW sense; this holds for SS and NS cells. Deviations of opposite senses are predicted for the upper surface. Fig.~\ref{Spot} shows that these predictions are correct. If the magnetic field is too small, viz., $| m  B_y |<| T_{\rm hyd}|$, so that the hydrodynamic torque always overcompensates the magnetic torque, more complex trajectories result.

In general, assuming that the speed in the $xy$ plane $v$, $T_{\rm hyd}$, $m$, and $B_y$ are constants, the position of a cell $(x,y)$ and the angle between its velocity and the horizontal magnetic field $\beta$, Fig.~\ref{model}, satisfy the following equations:
\begin{subequations}
\begin{align}
\frac{{\rm d}\beta}{{\rm d} t}=\,&\frac{T_{\rm hyd}-m\,B_y\,\sin\beta}{\xi} =\omega \big(1-p\, \sin\beta\big) \label{DGLa}\\
\frac{{\rm d}x}{{\rm d} t}=\,&-v \sin\beta  \   \label{DGLb}\\
\frac{{\rm d}y}{{\rm d} t}=\,&-v \cos\beta\ \label{DGLc}
\end{align}
\end{subequations}
where $\omega=T_{\rm hyd}/\xi$, $p=m B_y/T_{\rm hyd}$, with $\xi$ being the friction coefficient of a cell for rotation around the $z$-axis,  Fig.~\ref{model}(a). With initial conditions $\beta_0=\beta(0)$ and  $x(0)=y(0)=0$, we obtain
\begin{subequations}
\begin{align}
\tan \frac{\beta(t)}{2} =\,&p+\tilde{p} \,\tan\left\{ \frac{\tilde{p}\,\omega\, t}{2}- \tan^{-1} \left[\frac{p-\tan(\frac{\beta_0}{2})}{\tilde{p}}\right]\right\}    \label{beta}\\
x(t)=\,&\frac{v}{p\,\omega} \left(\beta(t)-\beta_0\right)-\frac{v\,t}{p}  \label{x}\\
y(t)=\,&\frac{v}{p\,\omega}\log \left( \frac{1-p^2}{1-p\,f(t)} \right), \label{y}
\end{align}
\end{subequations}
where 
\begin{equation}
\tilde{p}=\sqrt{1-p^2},
\end{equation}
\begin{equation}
\begin{aligned}
\label{f} f(t) = \,\sin\beta_0\, -\,&\tilde{p} \sin(\tilde{p}\,\omega\,t) \cos\beta_0 \\
+\,& \cos(\tilde{p}\,\omega\,t) \left(p-\sin\beta_0\right). 
\end{aligned}
\end{equation}

For $|p|>1$ (i.e. $|T_{\rm hyd}|<|m B_y|$) straight trajectories result, as expected, with fixed deviation $\beta=\sin^{-1}(p^{-1})$. For $|p|<1$, these equations predict loops whose shape, size, pitch and sense of rotation depend on $\omega$ and $p$. These loop swimmers should migrate along $x$, i.e. perpendicular to the horizontal field, at a rate that is essentially determined by $v/p$. Predicted trajectories for different values of $(p,\omega)$ are shown in Fig.~\ref{model}(b).

\subsection*{Trajectories and parameters}
We now analyse observed looped trajectories to obtain the parameters of our model. From an observed trajectory, we measure the time-dependent angular deviation, $\beta(t)$. This function is first fitted to Eq.~\ref{beta} to give $\beta_0 = \beta(t=0)$, $p$ and $\omega$. These values are then used in Eqs.~\ref{x} and \ref{y} to calculate the expected trajectory, using the average speed over the measured trajectory for $v$ in these expressions. 

An example prediction is shown in Fig.~ \ref{InterestingTracks}(c), where the inset shows the measured $\beta(t)$ (points) of the observed trajectory given in the main panel and the fitted function (continuous line). The fit parameters obtained, $(\beta_0, p, \omega)$, yield the predicted trajectory in the main pane (thin black curve), which shows all the important features (shape, size of loops, periodicity, migration predominantly along $x$) of the observed trajectory (thick red curve), and a reasonable quantitative fit up to and slightly beyond the first loop. 

In our model, loops occur when $|T_{\rm hyd}| > |mB_y|$, i.e. when the torque due to hydrodynamic interactions with the surface is larger than the magnetic torque due to the horizontal magnetic field. Two pieces of evidence suggest that this is indeed the case. First, we only observed loops when $|B_z| \gtrsim 1$~mT. We have already reported the observation from direct imaging that cells appear to swim closer to the surface at higher vertical fields. While the origin of this effect is unclear, it is reasonable to assume that cells swimming closer to the surface will experience a higher $T_{\rm hyd}$, consistent with the observation of loops only when $|B_z|$ is high enough.

Secondly, in a population of cells experiencing the same $B_z$, those cells with a lower $mB_y$, i.e. a smaller magnetic moment (because $B_y$ is constant for all cells) should be the ones that swim in loops. From fitting the 20 or so loop trajectories, we obtain an average value for  the ratio of magnetic to hydrodynamic torque, $| p | = 0.72\pm0.22$. Under the same field conditions, straight-swimming cells yield an average $p= 4.0 \pm 1.5$ (obtained in this case from $p=1/ \sin\beta$). Since $|B_z|$ is in the same range for both cases, we take $|T_{\rm hyd}|$ to be approximately constant. The observed values of $|p|$ then suggest that the magnitude of the magnetic moment $| m |$ of the loop-swimmers is $\lesssim 20\%$ of that of the straight swimmers. Since the magnetic moment of each cell depends not only on the number of magnetosomes but also their intracellular arrangement \citep{Faivre2010}, it is easy to envisage a distribution of $|m|$. Presumably, the loop swimmers come from the low-end tail of such a distribution.

To proceed further, we need a value for the average magnetic moment of {\it M. gryphiswaldense} cells. We know of two estimates of this quantity in the literature. Analysis of trajectories in rotating magnetic fields gives an upper bound of $m < 12 \times 10^{-16}$~A$\cdot$m$^2$ based on an estimated rotational friction coefficient $\xi$\citep{Erglis2007}. A  measurement using static light scattering \citep{Catalin} returned a fitted moment of $25 \times 10^{-16}$~A$\cdot$m$^2$. However, this work used a homogeneous cylinder to model the scattering of the cells, and returned a fitted cell length of $1.6~\mu$m, which is unrealistically short. In the closely related species \emph{M. magnetotacticum}, values in the range from 3.0 to $6.1\times 10^{-16}$~A$\cdot$m$^2$ were found \citep{Chemla1999, Bahaj1996}. 
Our measurements of the magnetic moment of non-motile {\em M. gryphiswaldense} cells using the DDM method \citep{MathiasDDM} described in Methods and Materials yields a value of $|m| = \left( 2.0 \pm 0.6 \right) \times 10^{-16}$~A$\cdot$m$^2$, comparable to previous values for {\it M. magnetotacticum}. 


Assuming that at high enough vertical field, $|B_z| \gtrsim 1$~mT, $\xi$ and $T_{\rm hyd}$ are approximately constant for all trajectories, we average over all trajectories to obtain $|\omega| = |T_{\rm hyd}/\xi|\approx 5.3 \pm 2.8$~rad/s. At $|B_y| = 1.5$~mT, straight trajectories gave us a fitted value of $| p | = mB_y/T_{\rm hyd} = 4.0\pm1.5$. These estimated values of $(|p|, |\omega|)$ together with the above measured value of $m$ then allow us to estimate $\xi \approx 14 \pm 8~$pN$\cdot$nm$\cdot$s. We compare this value to that of a prolate spheroid rotating around its minor axis $\xi_{s}=32\pi\eta a b^2 \epsilon^3 (2-\epsilon^2)/3(1-\epsilon^2)\left[(1+\epsilon^2)\log{\frac{1+\epsilon}{1-\epsilon}}-2\epsilon\right]$ \citep{Chwang1974}.  From fitting the diffusivity of non-motile cells in a magnetic field using a prolate ellipsoid model (see Materials and Methods), we found $a\approx 2.1$~$\mu$m and  $b\approx 0.6$~$\mu$m. Using $\eta=10^{-3}$~Pa$\cdot$s for water, we obtain $\xi_{s}=54$~pN$\cdot$nm$\cdot$s. 
The order of magnitude agreement with our estimate of $\xi= 14 \pm 8~$pN$\cdot$nm$\cdot$s from parameters fitted to observed trajectories is reassuring, especially since modelling the drag of a helix with a prolate ellipsoid in which the radius of the former and semi-minor axis of the latter are approximately equal will likely lead to a significant overestimation of the helix's rotational drag coefficient.

\section*{Discussion and Conclusions}

Our findings raise a number of intriguing questions for the biology of {\em M. gryphiswaldense}. In particular, since our results have unequivocally established the existence of fast and slow swimming modes for each cell, we may enquire into the molecular {\it mechanism} for such mode switching, and into its {\it function} in the cells' natural environment. Turning first to the functional question, we suggest that this phenotype controls the ability of a swimming cell to explore its environment in directions {\it perpendicular} to the magnetic field, $\mathbf{B}$. A self-propelled particle with finite magnetic moment in an external magnetic field will travel {\it on average} parallel to $\mathbf{B}$; but its rotational Brownian motion will cause directional fluctuations, so that the particle will explore dimensions orthogonal to the field lines. The larger the particle's propulsion velocity is, the further the particle will move away from the original field line before the magnetic field restores its orientation along $\mathbf{B}$. It is therefore possible that a cell uses these two swimming modes to `tune' how much it wanders away from swimming along $\mathbf{B}$, e.g. as a function of how close it is to its preferred microaerophilic environment. A second suggestion is that {\it M. gryphiswaldense} somehow uses its two speeds directly in magnetoaerotaxis analogous to how {\it E. coli} can use speed modulation to navigate temperature gradients (thermotaxis) \citep{SalmanThermo}. 

A final possibility is based on the result that the optimal `sensing time' required for detecting a (logarithmic) concentration gradient $\nabla \ln c$ is proportional to $(v\nabla \ln c)^{-2/3}$ (Eq.~58 in \citep{Purcell}). Thus, switching from a fast to a slow swimming mode such that $v_{\rm fast}/v_{\rm slow} \approx 3$ allows a cell to sample a $|\nabla \ln c|$ that is three times smaller, so that {\em M. gryphiswaldense} perhaps utilises this mechanism to optimise oxygen sensing. 

We observed that cells switched between fast and slow swimming modes rapidly. Indeed, the switching time is below our current experimental resolution. Since we used running averages of 20 frames at 100 frames per second to construct trajectories, we can give an upper bound: switching between swimming modes in {\it M. gryphiswaldense} occurs in $\lesssim 0.2$~s. Our theoretical model supports the suggestion that the switching is between using leading and trailing flagella. We may therefore, secondly, enquire into the molecular mechanism used by a cell to achieve such fast and coordinated changes in the motors situated at its two opposite poles. 

While only the most preliminary mechanistic suggestions could be offered at this stage, e.g. the use of a molecular `clutch' (as used in {\em Bacillus subtilise} to switch motility `on' and `off' rapidly and reversibly \citep{Blair2008}), we can certainly set our upper bound for the switching time in the context of intracellular diffusion. Flagellar motion in {\it E. coli} is controlled by the diffusion of proteins such as CheY and CheZ, whose diffusion coefficients in the cytoplasm are in the range 5-10~$\mu$m$^2$s$^{-1}$ \citep{BrayDiffusion}. To diffuse from one pole to another, a distance of $L \approx 4~\mu$m, takes $t \approx L^2/6D \approx 0.3$~s, which is consistent with our upper bound, and with recent work concluding that diffusion of CheY suffices to explain sub-second coordinated switching between CCW and CW flagellar rotation in this organism \citep{Terasawa2011}. Presumably, intracellular diffusion of relevant analogous `proteins' also lies behind the coordinated switching between leading and trailing flagella in our organism. Interestingly, various ionic salts are known to disrupt the coordinated switching of polar bundles between CCW and CW rotation in {\em S. voluntas} \citep{Krieg1967}. It would be interesting to perform similar experiments in {\it M. gryphiswaldense}.

Switching between swimming modes, and indeed between NS and SS behaviour, require the coordination of flagellar motors.  Again, only the most preliminary suggestion could be made based on analogy with other organisms.  Very recent work shows that the intracellular diffusion of phosphorylated CheY, long known to be a key molecule in chemotaxis in {\it E. coli}, is sufficient to explain the coordinated switching between CCW and CW rotation of flagellar motors in this species on a sub-second time scale. The similarity in cell size and switching time scales between {\em E. coli} and {\em M. gryphiswaldense} suggest once the relevant protein(s) have been identified in the latter, their intracellular diffusion should again suffice to explain flagellar coordination. 

\section*{acknowledgments}
Funding came from the Swiss National Science Foundation (PBFRP2-127867), FP7-PEOPLE (PIIF-GA-2010-276190), and the UK's EPSRC  (EP/D071070/1, EP/E030173, and EP/I004262/1).

%
%
%
%

\clearpage
\section*{Figure Legends}

\subsection*{Figure~\ref{fig:D_vs_B}.}
Effective translational diffusion coefficients versus the amplitude of the magnetic field in the direction perpendicular, $D_{\perp}$, and parallel, $D_{\parallel}$, to the applied magnetic field. Error bars are standard deviations, obtained from averaging the measured diffusion coefficients over the range of spatial frequency $1\leq q \leq 2.3~\mu$m$^{-1}$, and representative of a slight $q$-dependency -- a measure of systematic error. Lines are simultaneous fits (global fitting) to the experimental data using Eqs. 1 and 2 with $m$, $D_a$, and $D_b$ linked for both fits. We obtained $m=2.0 \pm 0.6 \times 10^{-16}~\text{A}\cdot\text{m}^2$, $D_a=0.244 \pm 0.003~\mu\text{m}^2$/s, and $D_b=0.192 \pm 0.002~\mu\text{m}^2$/s.

\subsection*{Figure~\ref{MicroCell}.}
A schematic of our observation chamber, defining axes, and showing how a magnetic field with finite $y$ and $z$ components separates NS and SS cells.

\subsection*{Figure~\ref{Polar_Bulk}.}
a) Swimming velocity distribution (color coded) obtained by tracking {\it M. gryphiswaldense} in the bulk at a horizontal applied field of $B_y=-1.5$ mT. b) and c) Trajectories of SS and NS cells, respectively, with color-coded swimming speed, illustrating mode switching: fast-stop-fast in b) and fast-slow in c).

\subsection*{Figure~\ref{Spot}.}
Swimming velocity distribution measured close to the top (a and b) and the bottom wall (c and d) with horizontal field $B_y=-1.5$~mT. The vertical field was $B_z=0.86$~mT for a) and c) and  $B_z=-0.86$~mT for b) and d).

\subsection*{Figure~\ref{FieldDependency}.}
Swimming velocity distribution measured close to the bottom wall for vertical field strength  $B_z$: $0.0$, $-0.12$, $-0.43$, $-0.64$~mT (from a) to d)) and constant horizontal field $B_y=-1.5$~mT. (e) Swimming angle $\beta$ (plotted as $\sin\beta$) close to the bottom wall for slow (black triangles) and fast (red squares) north seekers. The points and error bars correspond to the mean values and full-width-half-maxima of the peaks, respectively. The lines are a guides to the eye.

\subsection*{Figure~\ref{InterestingTracks}.}
Trajectories of individual cells swimming in the indicated directions in a horizontal field of $B_y=-1.5$ mT. The color code corresponds to the absolute swimming speed. Trajectories a) and b) ($B_z = + 0.86$~mT) were recorded close to the top and c) ($B_z = -1.0$~mT) close to the bottom of the observation chamber. The line in c) shows a calculated trajectory with the best-fit parameters $p=-0.604$, $\omega=-6.88$~rad.s$^{-1}$  and $v = 52~\mu$ms$^{-1}$. Inset: calculated (line) and measured $\beta(t)$ (points).

\begin{figure}
\begin{center}
\includegraphics[width=3.25in]{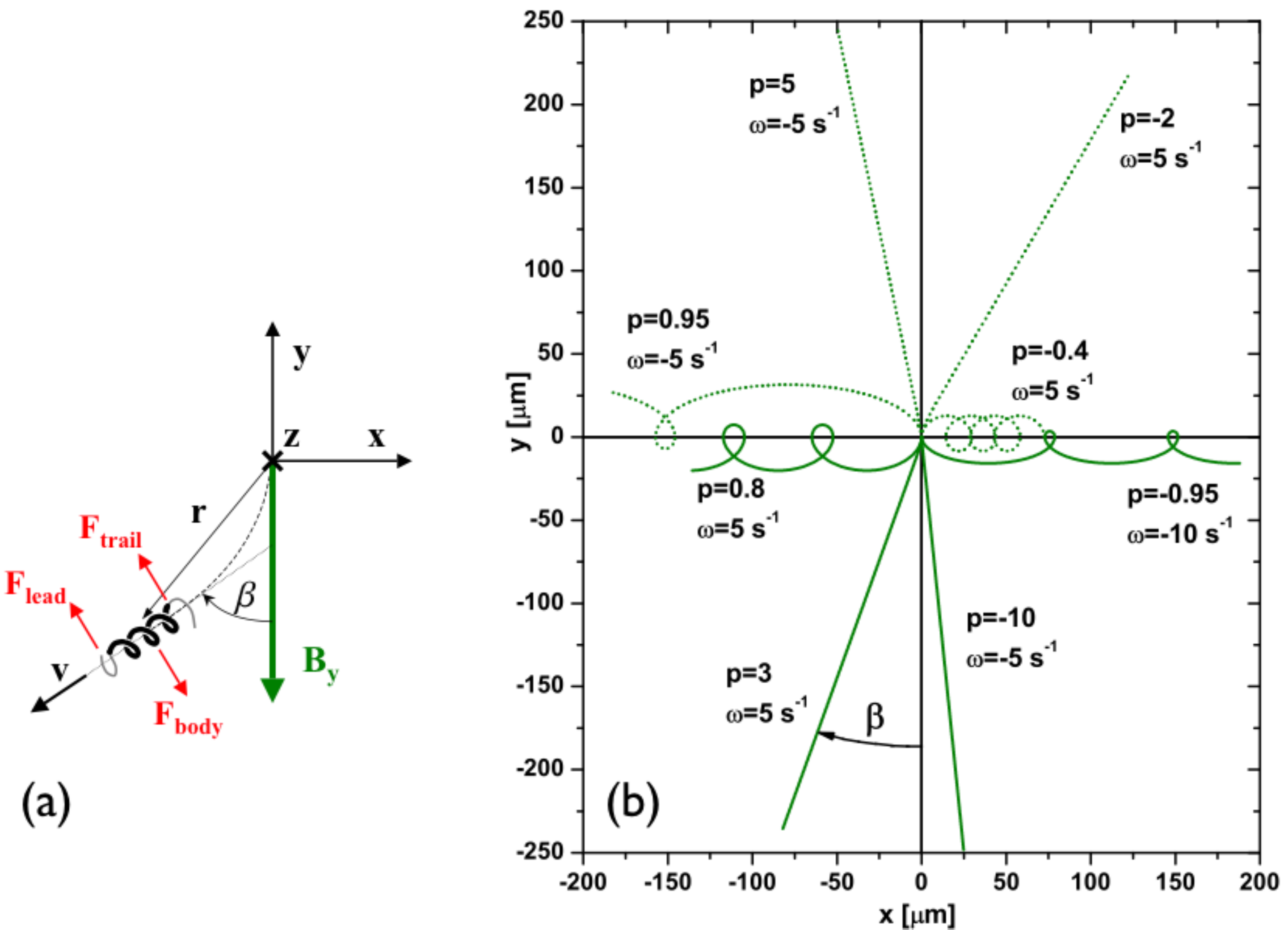}
\caption{}
\label{model}
\end{center}
\end{figure}

\subsection*{Figure~\ref{model}.}
(a) Schematic of a NS swimming along the bottom of the observation chamber. Seen from behind, the right-handed helical body rotates CW and the flagella CCW. Red arrows show the directions of the surface-mediated hydrodynamic forces on the body and flagella when they are rotating in these respective senses. (b) Calculated trajectories for different parameters $p$ and $\omega$ in a magnetic field $B_y=-1.5$~mT. Solid lines represent NS ($m<0$) and dotted lines SS ($m>0$) cells. All trajectories are calculated over 5 s for  $v=50$~$\mu$m/s  and with initial conditions $x(0)=y(0)=0$.


\clearpage

\clearpage

\clearpage

\clearpage

\clearpage

\clearpage

\clearpage

\end{document}